\documentclass[aps,prl,twocolumn,amsmath,amssymb,nofootinbib,superscriptaddress,floatfix]{revtex4-1}

\usepackage{amsmath}
\usepackage{amssymb}
\usepackage{amsthm}
\usepackage{mathtools}
\usepackage{mathdots}
\usepackage{amsfonts}
\usepackage{bbm}
\usepackage{xr}
\usepackage{bbold}
\usepackage{newpxtext}
\usepackage{epsfig,color,dsfont,upgreek,physics}
\usepackage{mathrsfs}
\usepackage{multirow}

\usepackage{graphicx}
\usepackage{dcolumn} 
\usepackage{bm} 
\usepackage{float} 
\usepackage[driverfallback=dvipdfm]{hyperref} 
\usepackage{longtable}
\usepackage{ulem}   
\allowdisplaybreaks

\newcommand{\bs}[1]{{\boldsymbol{#1}}}

\newcommand{\sgn}{\mathcal{\text{sgn}}}

\renewcommand\[{\begin{equation}}
\renewcommand\]{\end{equation}}

\begin{document}

\title{Triplet Pair-Density Wave Superconductivity on the \(\pi\)-flux Square Lattice}

\author{Daniel Shaffer}
\affiliation
{
Department  of  Physics,  Emory  University,  400 Dowman Drive, Atlanta,  GA  30322,  USA
}

\author{Luiz H. Santos}
\affiliation
{
Department  of  Physics,  Emory  University,  400 Dowman Drive, Atlanta,  GA  30322,  USA
}

\begin{abstract}

Pair-density waves (PDW) are superconducting states that spontaneously break translation symmetry in systems with time-reversal symmetry (TRS). Evidence for PDW has been seen in several recent experiments, as well as in the pseudogap regime in cuprates.  Theoretical understanding of PDW has been largely restricted to phenomenological and numerical studies, while microscopic theories typically require strong-coupling or fine-tuning. In this work, we provide a novel symmetry-based mechanism under which PDW emerges as a weak coupling instability of a 2D TRS metal. Combining mean-field and renormalization group analyses, we identify a weak-coupling instability towards a triplet PDW realized in the \(\pi\)-flux square lattice model with on-site repulsion and moderate nearest-neighbor attraction when the Fermi level crosses Van Hove singularities at 1/4 and 3/4 fillings. This PDW is protected by the magnetic translation symmetries characteristic of Hofstadter systems, of which the \(\pi\)-flux lattice is a special time-reversal symmetric case.

\end{abstract}

\date{\today}

\maketitle

\noindent

\noindent
\textit{Introduction--} Several recent experimental works \cite{HamidianDavis16, Edkins19, LiuDavis20,Chen21, Liu22, AishwaryaFradkinMadhavan22, GuPaglioneDavis22} have reported an exotic form of superconductivity called a pair-density wave (PDW). There is also longstanding indirect evidence for PDW as a potential explanation for the coexistence of superconductivity with broken translation orders \cite{FradkinKivelsonTranquada15}. PDW states are characterized by a non-uniform order parameter that, unlike Fulde–Ferrell–Larkin–Ovchinnikov (FFLO) states \cite{FF,LO}, is spontaneously formed in a normal state that preserves time-reversal symmetry (TRS). While phenomenological theories have been profusely used in the description of coarse-grained properties of PDWs (see Ref. \cite{Agterberg20} and references therein for a comprehensive review), 
microscopic theories have been mostly restricted to one dimensional systems \cite{Zachar01, RobinsonTsvelik12, JaefariFradkin12, VenderleyEunAhKim19, ZhangVishwanath22, HuangKivelson22}, and systems in the strong-coupling regime \cite{Loder11, BergFradkinKivelson09II, KangTesanovich11, Soto-GarrindoFradkin15, WuRaghu22}. Very few models have a weak-coupling PDW instability \cite{Lee14, Soto-GarridoFradkin14}, including more recent models invoking renormalization group (RG) methods \cite{HsuEunAhKim17, Wu22, WuWuWu22, ShafferBurnellFernandes22}, but in all these models some degree of fine-tuning is required as the weak-coupling instability is not symmetry-protected in those cases.

\begin{figure*}[t]
\centering
\includegraphics[width=0.99\textwidth]{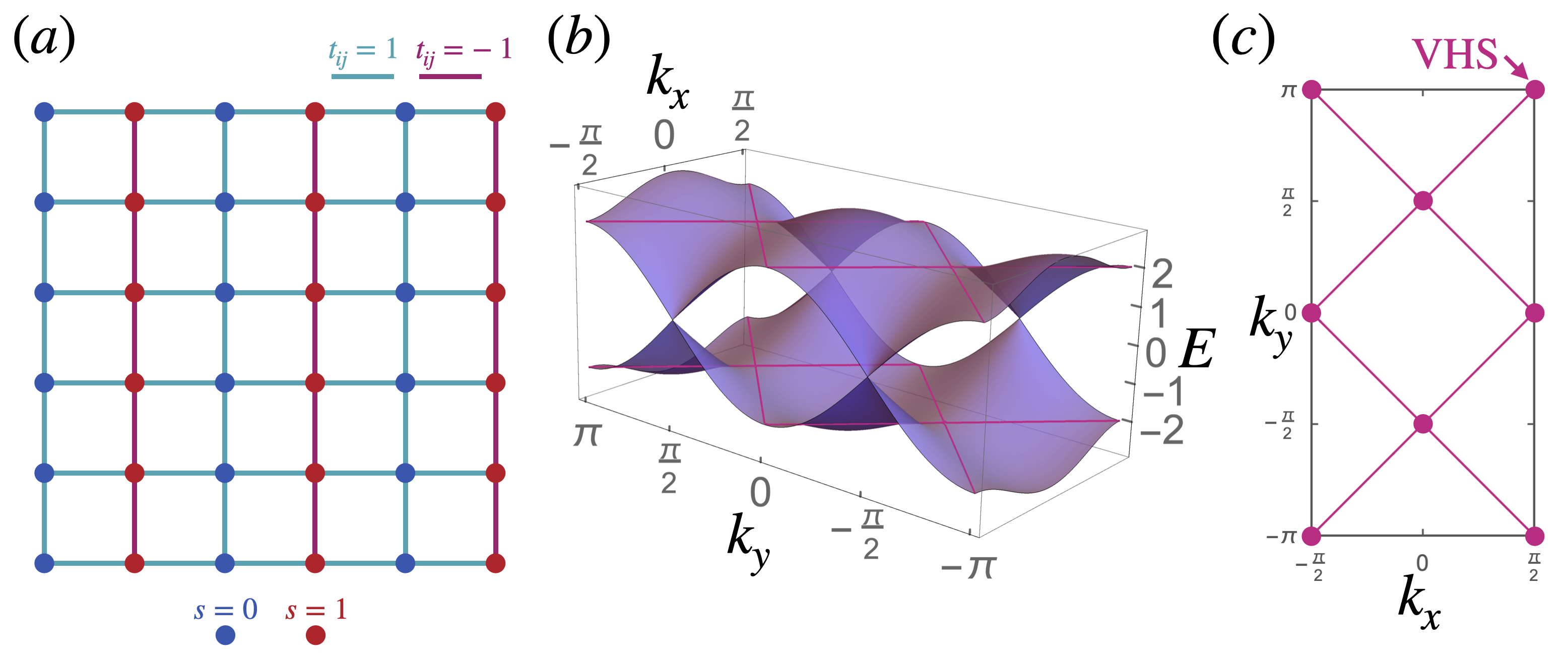}
\caption{(a) The square \(\pi\)-flux lattice. Blue and red sites correspond to the two sublattice site \(s=0\) and \(1\) respectively. The bonds represent nearest neighbor hopping amplitudes \(t_{ij}=t(-1)^{(r_y-r_y')r_x}\) that are either \(1\) or \(-1\) (blue and red respectively). (b) The two bands of the \(\pi\)-flux lattice dispersion defined on the MBZ. There are two Dirac nodes on the MBZ boundary at \(\pm(\pi/2,\pi/2)\). The red energy contour corresponds to the VHS, with the corresponding Fermi surface and location of the VHSs in the MBZ shown in (c).
}
\label{fig:Model}
\end{figure*}

In this work we show that a symmetry-protected PDW state can be realized at weak-coupling without fine-tuning in the interacting \(\pi\)-flux square lattice model, a special case of the Hofstadter model \cite{Hofstadter76}, when the Fermi level is set to 1/4 or 3/4 fillings corresponding to the Van Hove singularities (VHSs) (see Fig. \ref{fig:Model}).
The realization of the Hofstadter bands \cite{Hofstadter76} in moiré lattices \cite{Dean13,Ponomarenko13,Hunt13, Saito21, YuKivelsonFeldman22} has boosted interest in the phenomenon of electronic pairing in fractal bands  \cite{ZhaiOktel10,SohalFradkin20,ShafferWangSantos21}.
As recently shown by the authors \cite{ShafferWangSantos21}, the non-trivial organization of the single particle states due to magnetic translation symmetry (MTS) allows for Cooper pairs with finite center-of-mass momentum, due to correlations between pairs of electrons across self-similar patches of the Fermi surface. Thus, Hofstadter bands, once the playground of quantum Hall physics, emerge as a new arena to explore exotic superconductivity with broken translation symmetry.

The \(\pi\)-flux lattice is a particularly special case as unlike other Hofstadter systems it possesses TRS, so that the finite-momentum pairing in this system in fact realizes a PDW state. The focus of earlier studies, however, has been  the onset of Dirac fermions at half-filling, notably (but not exclusively) in the context of flux phases in cuprates \cite{AffleckMarston88,Anderson89,Lieb94}, quantum Hall transitions \cite{ludwig1994integer} and electron fractionalization \cite{Wen-Wilczek-Zee1989, neupert2011fractional,neupert2011fractionalliquids} in 2D lattices. Here our attention is shifted away from half-filling and, instead, focused on the less explored fermionic states near $1/4$ and $3/4$ fillings, which realize perfectly nested Fermi surfaces with a novel structure of VHSs. 

The perfect nesting condition together with diverging density of states (DOS) near VHSs that arise due to the MTS, enhance spin fluctuations that open the door to unconventional electronic pairing due to logarithmically divergent susceptibilities \cite{Scalapino12}. Exploring this possibility in RG, we had earlier identified a new weak coupling instability towards a uniform singlet d-wave like superconductor driven by on-site repulsive Hubbard interaction \cite{ShafferWangSantos22}. The main result of this work is the identification of a new instability towards a triplet PDW ground state when moderate nearest neighbor attractive interactions are added. We establish the existence of the PDW state using mean-field theory and confirm its stability with RG methods that rule out alternative competing phases (SC, CDW and SDW). Crucially, the weak-coupling triplet PDW instability is protected by the magnetic translation symmetries of the \(\pi\)-flux lattice, thus representing a new class of unconventional superconductivity.

\textit{The \(\pi\)-flux Lattice Model --} 
The \(\pi\)-flux square lattice Hamiltonian is
\[\label{H0}
H_0 =-t\sum_{\langle \mathbf{r} \mathbf{r}' \rangle \sigma} \, (-1)^{(r_y-r_{y}')r_x} c_{\mathbf{r}\sigma}^\dagger c_{\mathbf{r}' \sigma} + h.c.
-\mu \sum_{\mathbf{r} \sigma} c_{\mathbf{r}\sigma}^\dagger c_{\mathbf{r} \sigma}\]
where \(c_{\mathbf{r}\sigma}\) are fermion annihilation operators at site \(\mathbf{r}\) of the square lattice with spin \(\sigma=\uparrow,\downarrow\), \(t\) is the nearest-neighbor hopping amplitude that we set to \(1\), and \(\mu\) is the chemical potential. We take \(r_x,\, r_y\) to be integer multiples of the lattice constant \(a=1\), and consider \(\mu=\pm 2t\) corresponding to energy of the VHSs. See Fig. \ref{fig:Model} (a). 

The symmetries of the \(\pi\)-flux model will play an important role in our analysis below. \(H_0\) possesses TRS and inversion symmetry, as well as magnetic translation symmetries \(\hat{T}_y=T_y\) and \(\hat{T}_x=(-1)^{r_y}T_x\) forming the magnetic translation group (MTG), where \(T_x\) and \(T_y\) are ordinary lattice translations. In particular, the translation \(T_x\) is broken by \(H_0\), so that there are two sites per unit cell along the \(x\) direction which we label with a sublattice index \(s=0,1\) as shown in Fig. \ref{fig:Model} (a). In addition, \(H_0\) has a four-fold magnetic rotation symmetry \(\hat{C}_4\) which is a combination of the regular \(C_4\) symmetry with a gauge transformation that correspondingly rotates the magnetic vector potential; see SM \cite{SM} for the explicit form (other crystalline symmetries of the square lattice have similar magnetic versions).

\(H_0\) is partially diagonalized by the momentum-space operators
\(c_{\mathbf{k}s\sigma}=\frac{1}{\sqrt{N}}\sum_{\mathbf{R}}e^{-i\mathbf{k}\cdot (s\hat{\mathbf{x}}+\mathbf{R})}c_{s\hat{\mathbf{x}}+\mathbf{R},\sigma}\)
where \(\mathbf{R}\) defines the lattice with the extended unit cell and \(N\) is the number of extended sites. The momentum \(\mathbf{k}\) is thus restricted to the magnetic Brillouin zone (MBZ) with \(k_y\in (-\pi,\pi]\) and \(k_x\in [-\pi/2,\pi/2]\), such that the original Brillouin zone (BZ) is folded along the \(k_x\) direction (see Fig. \ref{fig:Model} (b-c)). In momentum space, the single-body Hamiltonian can be written as \(H_0=\sum_{\mathbf{k}\sigma}\mathbf{c}_{\mathbf{k}\sigma}^\dagger\mathcal{H}_0(\mathbf{k})\mathbf{c}_{\mathbf{k}\sigma}\) where \(\mathbf{c}_{\mathbf{k}\sigma}=(c_{\mathbf{k},0,\sigma},c_{\mathbf{k},1,\sigma})^T\) and \(\mathcal{H}_{0}(\bs{k}) = 2\,t\,\cos{k_x}\,\tau^x -2\,t\,\cos{k_y}\,\tau^z -\mu\), with \(\tau^j\) being Pauli matrices acting on sublattice indices. \(\mathcal{H}_0(\mathbf{k})\) is diagonalized by operators
\[d_{\mathbf{k}\alpha\sigma}=\sum_s\frac{\alpha^s}{\sqrt2}\sqrt{1+\alpha (-1)^s \frac{\cos k_y}{E(\mathbf{k})}} c_{\mathbf{k}s\sigma}\label{d}\]
where \(E(\mathbf{k})=\sqrt{\cos^2 k_x+\cos^2 k_y}\) and \(\alpha=\pm1\) labels the Hofstadter bands with energies
\(\epsilon_\alpha(\mathbf{p})=2t\alpha E(\mathbf{k})-\mu\).
The two bands meet at zero energy at two Dirac nodes located at the MBZ boundary at \(\mathbf{k}=\pm(\pi/2,\pi/2)\), see Fig. \ref{fig:Model} (b).

In this work we investigate electronic states near $1/4$ and $3/4$ fillings at which the bottom and top bands, respectively, have \(4\) VHSs, as shown in Fig. \ref{fig:Model} (c). 
Crucially, we will show that the diverging density of states at the VHSs gives rise to various pairing instabilities, including a triplet PDW state.

\textit{Extended Hubbard Interactions and Irreducible Representations - } To investigate interaction effects, we add to Eq. (\ref{H0}) an extended Hubbard interaction:
\[H_I=U\sum_{\mathbf{r}}
n_{\mathbf{r}\uparrow}\,n_{\mathbf{r} \downarrow}+V\sum_{\langle \mathbf{r} \mathbf{r}' \rangle \sigma\sigma'}n_{\mathbf{r}\sigma}\,n_{\mathbf{r'}\sigma'}
\label{HI}\,,\]
where \(n_{\mathbf{r}\sigma}=c_{\mathbf{r}\sigma}^\dagger c_{\mathbf{r} \sigma}\) is the electron density operator, and \(U\) and \(V\) are the on-site and nearest-neighbor density-density interaction strengths, respectively. We consider both positive (repulsive) and negative (attractive) values of \(U\) and \(V\). Following the approach in \cite{ShafferWangSantos22}, we project the interactions onto the Hofstadter band \(\alpha\) crossing the chemical potential tuned to the VHS. This is done by replacing the \(c_{\mathbf{r} \sigma}\) operators in Eq. (\ref{HI}) with the band basis \(d_{\mathbf{k}\alpha}\) operators  from Eq. (\ref{d}), and restricting the \(\alpha\) index to a single value.

We further decompose the interactions into pairing channels corresponding to the different irreducible representations (irreps) of the symmetries of the \(\pi\)-flux lattice. The relevant irreps of the MTG symmetries for arbitrary rational flux have been classified in \cite{ShafferWangSantos21}. The key observation for constructing the irreps is that the operator \(\hat{T}_x\hat{T}_y\hat{T}_x^{-1}\hat{T}_y^{-1}\) corresponding to a loop operator going around a single unit cell of the square lattice is a \(U(1)\) transformation under which the electrons pick up an Aharonov-Bohm phase equal to the encircled flux, a phase of \(\pi\) on the \(\pi\)-flux lattice. A \emph{pair} of electrons, on the other hand, thus picks up a trivial phase, such that the action of \(\hat{T}_x\) and \(\hat{T}_y\) commutes on the level of the pairing gap function. As a result, we can classify the gap functions according to it being even or odd under the MTG symmetries, with all four combinations being possible. These correspond to four one-dimensional irreps of the MTG that we label \((\ell_x,\ell_y)\) with \(\ell_j=0,1\) corresponding to Cooper pairs with momenta \(\pi(\ell_x,\ell_y)\) and picking up a phase of \((-1)^{\ell_j}\) under \(\hat{T}_j\). The uniform SC state corresponds to \(\ell_x=\ell_y=0\), and the other three irreps are PDW orders.

In addition, the electrons can form either singlet or triplet pairs (the irreps of spin rotation symmetry, unbroken by the spin-conserving interactions Eq. (\ref{HI})). We thus label the gap functions as \(\Delta^{(\ell_x\ell_y\nu)}\), with \(\nu=0\) corresponding to singlet and \(\nu=x,y,z\) corresponding to the three triplet components.
Finally, when \(\ell_x=\ell_y\), the Cooper pair momentum is invariant under \(\hat{C}_4\) and the gap function can thus be even or odd under \(\hat{C}_4\), which we refer to as \(s\)-wave or \(d\)-wave pairing respectively and label the corresponding gap functions as \(\Delta^{(\ell_x\ell_y\nu;s)}\) and \(\Delta^{(\ell_x\ell_y\nu;d)}\).
When \(\ell_x\neq \ell_y\), since \(\hat{C}_4\) interchanges the action of \(\hat{T}_x\) and \(\hat{T}_y\), the gap functions necessarily break the \(\hat{C}_4\) symmetry, implying that there are at least two degenerate ground states of the system. In particular, \(\Delta^{(01\nu)}\) and \(\Delta^{(10\nu)}\), which correspond to pairs with total momenta \(\bar{\mathbf{Q}}=(\pi,0)\) and \(\mathbf{Q}=(0,\pi)\) respectively, are mapped to each other under \(\hat{C}_4\). This means that although they form two 1D irreps of the MTG taken separately \cite{ShafferWangSantos21}, taken together they form two components of a 2D irrep of the MTG combined with \(\hat{C}_4\).

The crystalline symmetry relations above determine the space group irreps describing the PDW order parameters \cite{AgterbergSigrist09}, which are direct products of the irrep of the little point group (the subgroup of the point group that keeps the total momentum of the pair fixed) and the star of the total momentum of the pair (i.e. the collection of all momenta mapped to each other by the point group symmetries). All the irreps and representative gap functions are summarized in Table \ref{table:irreps}, with the relevant little point groups being the trivial group for channels with the star \(\mathbf{Q}^*=\{\mathbf{Q},\bar{\mathbf{Q}}\}\) and \(C_{4h}\) otherwise (the smallest point group containing inversion and \(\hat{C}_4\) symmetry).

Having identified the irreps, the interactions projected onto the band \(\alpha\) can be decomposed into the corresponding channels as
\begin{widetext}
\[H^{(\ell_x\ell_y\nu)}_{int,\alpha}=\frac{1}{2}\sum_{\mathbf{k p}\sigma_1\sigma_1'\sigma_2\sigma_2'} g^{(\ell_x\ell_y\nu)}(\mathbf{p;k})(\sigma^\nu i\sigma^y)^*_{\sigma_1\sigma_1'}(\sigma^\nu i\sigma^y)_{\sigma_2\sigma_2'}d^\dagger_{\mathbf{p}+\ell_x\mathbf{Q},\alpha\sigma_1}d^\dagger_{-\mathbf{p}\alpha\sigma_1'}d_{-\mathbf{k}\alpha\sigma_2}d_{\mathbf{k}+\ell_x\mathbf{ Q},\alpha\sigma'_2}\label{HIproj}\]
\end{widetext}
where
\[g^{(\ell_x\ell_y\nu)}(\mathbf{p;k})=\sum_{m}g^{(\ell_x\ell_y\nu)}_{m}\Phi^{(\ell_x\ell_y\nu)}_m(\mathbf{p})\Phi^{(\ell_x\ell_y\nu)}_m(\mathbf{k})
\label{eq:g functions}
\]
with \(g^{(\ell_x\ell_y\nu)}_{m}\) being the coupling \emph{constants} with \(m=0,1,\dots\) labeling terms belonging to the same irrep and \(\Phi^{(\ell_x\ell_y\nu)}_m(\mathbf{p})\) are basis functions; \(g^{(\ell_x\ell_y\nu)}_{m}\) and \(\Phi^{(\ell_x\ell_y\nu)}_m(\mathbf{p})\) obtained from the extended Hubbard interactions are summarized in Table \ref{table:irreps}. Note that the projections in this case do not depend on the band index \(\alpha\), which we therefore henceforth omit.

Having obtained the projected pairing interactions Eq. (\ref{HIproj}), we analyse the resulting instabilities using 
two complementary methods: a mean-field analysis, and a patch RG calculation following the method described in \cite{ShafferWangSantos22}.

\begin{table}[htp]
\begin{center}
\begin{tabular}{ c c c c }
\hline\hline\noalign{\smallskip}
Sp. Grp. & \multirow{2}{*}{\((\ell_x\ell_y\nu)\)} & \multirow{2}{*}{\(\Phi^{(\ell_x\ell_y\nu)}_m(\mathbf{k})\)} & \multirow{2}{*}{\(g^{(\ell_x\ell_y\nu)}_{m}\)} \\
Irrep\\
\noalign{\smallskip}\hline\hline\noalign{\smallskip}
\multirow{2}{*}{\(\boxed{A_{g}}\)} &\multirow{2}{*}{\((00s;s)\)} &  1 & \(U/2\)\\
& &   \((\cos^2k_y+\cos^2k_x)/E(\mathbf{k})\) & \( V\)\\
\noalign{\smallskip}\hline\noalign{\smallskip}
\(\boxed{B_{g}}\) &\((00s;d)\) &  \((\cos^2k_y-\cos^2k_x)/E(\mathbf{k})\)  & \(V\) \\
\noalign{\smallskip}\hline\noalign{\smallskip}
 \multirow{4}{*}{\(\mathbf{Q}^{*}_s\)} & \multirow{2}{*}{\((01s)\)} & \(\cos k_y/E(\mathbf{k})\) & \(U/2\)\\
&  &  \(\cos k_y\) & \(V\)\\
\multirow{2}{*}{} & \multirow{2}{*}{\((10s)\)} & \(\cos k_x/E(\mathbf{k})\) & \(U/2\)\\
&  & \(\cos k_x\) & \( V\)\\
\noalign{\smallskip}\hline\noalign{\smallskip}
 \multirow{2}{*}{\(E_g\)} & \multirow{2}{*}{\((10s)\)} & \(\sin k_y\cos k_x/E(\mathbf{k})\) & \(V\)\\
&  & \(\sin k_x\cos k_y/E(\mathbf{k})\) & \(V\)\\
\noalign{\smallskip}\hline\noalign{\smallskip}
 \multirow{2}{*}{\(E_{u}\)} & \multirow{2}{*}{\((00t)\)} & \(\sin k_y\cos k_y/E(\mathbf{k})\) & \(V\)\\
& &  \(\sin k_x\cos k_x/E(\mathbf{k})\) & \(V\)\\
\noalign{\smallskip}\hline\noalign{\smallskip}
\multirow{2}{*}{\(\boxed{\mathbf{Q}^{*}_t}\)} &\((01t)\)  & \(\sin k_y\) & \(V\)\\
  & \((10t)\) & \(\sin k_x\) & \(V\)\\
  \noalign{\smallskip}\hline\noalign{\smallskip}
 \(A_{u}\) & \((11t)\) & \(\cos k_x\cos k_y/E(\mathbf{k})\) & \(V\)\\
\noalign{\smallskip}
\hline\hline
\end{tabular}
\caption{Basis functions \(\Phi^{(\ell_x\ell_y\nu)}_m(\mathbf{k})\) for gap functions and interactions arising in the extended Hubbard model belonging to various irreducible representations. The first column shows the space group irrep composed of the little point group irrep times the star of the momentum of the order. If only the point group is shown, the momentum star is trivial. The non-trivial momentum star is \(\mathbf{Q}^*=\{(0,\pi),(\pi,0)\}\), for which the little point group is trivial. We further label the momentum star \(\mathbf{Q}^*_s\) and \(\mathbf{Q}^*_t\) to distinguish the singlet and triplet channels. The second column shows the \((\ell_x\ell_y\nu)\) labels. \(\ell_j=0,1\) correspond to gap functions even and odd under \(\hat{T}_j\) respectively, which determines the MTG irrep. \(\nu=s,t\) labels singlet and triplet channels. \((00s;s)\) and \((00s;d)\) indicate the \(s\)- and \(d\)-wave channels respectively. The third column shows \(\Phi^{(\ell_x\ell_y\nu)}_m(\mathbf{k})\), with multiple rows in a single channel corresponding to different values of \(m\), meaning either several functions belong to the same 1D irrep or they from two components of the 2D irreps. The last column shows the coupling constants that give rise to the corresponding channels. The channels that we find to be leading in our model are boxed.
}
\label{table:irreps}
\end{center} 
\end{table}

\textit{Mean-Field Analysis --}
The mean field pairing Hamiltonian reads
\[H_\Delta=\sum_{\ell_x\nu\mathbf{p}} \Delta^{(\ell_x\nu)}(\mathbf{p})(\sigma^\nu i\sigma^y)_{\sigma\sigma'}d^\dagger_{\mathbf{p}+\ell_x\mathbf{Q},\sigma}d^\dagger_{-\mathbf{p}\sigma'}+h.c.\]
with \(\Delta^{(\ell_x\nu)}(\mathbf{p})\) being the gap functions, which, near the phase transition at the critical temperature \(T_c\), satisfy the linearized gap equation
\[\Delta^{(\ell_x\nu)}(\mathbf{p})=-\sum_{\ell_y,\mathbf{k}} g^{(\ell_x\ell_y\nu)}(\mathbf{p};\mathbf{k})\Pi(\mathbf{k})\Delta^{(\ell_x\nu)}(\mathbf{k})\label{LinGapEq}\]
where \(\Pi(\mathbf{k})=\tanh\left(\frac{\epsilon_\alpha(\mathbf{k})}{2T}\right)/\epsilon_\alpha(\mathbf{k})\) is the pairing susceptibility. Observe that the susceptibility has a logarithmic divergence and is independent of \(\ell_x\), which is a direct consequence of the \(\hat{T}_x\) symmetry that identifies states at \(\mathbf{k}\) and \(\mathbf{k+Q}\). We emphasize that this means that the weak-coupling PDW instability is symmetry-protected in the \(\pi\)-flux model along with the usual uniform SC instability.

\begin{figure}[t]
\centering
\includegraphics[width=0.49\textwidth]{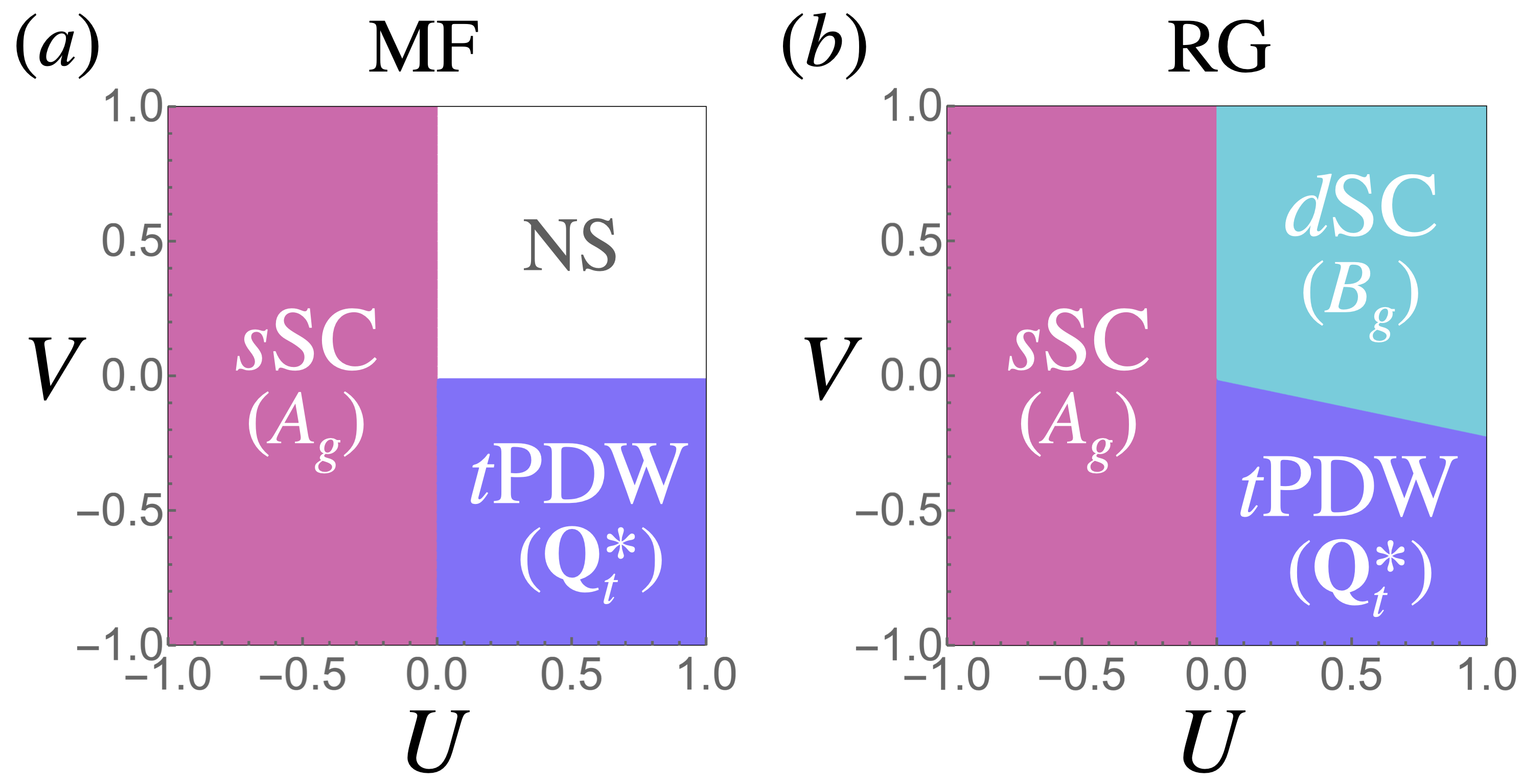}
\caption{Phase diagrams for the square \(\pi\)-flux lattice with extended Hubbard interactions in the parameter space of \(U\) (on-site) and \(V\) (nearest-neighbor) density-density interaction strengths (units are arbitrary), obtained using (a) mean-field and (b) patch RG calculations. In both cases on-site attraction (\(U<0\)) results in a uniform \(s\)-wave SC belonging to the \(A_g\) irrep in Table \ref{table:irreps}, while nearest-neighbor attraction with on-site repulsion (\(V<0,\, U>0\)) results in a triplet PDW in the \(\mathbf{Q}^*_t\) irrep in Table \ref{table:irreps}. In the latter case nodal unidirectional and fully gapped bidirectional PDW ground states are degenerate. The difference in the RG analysis is that a \(d\)-wave uniform SC phase in the \(B_g\) irrep is realized even when all interactions are repulsive, as well as for \(-0.21U<V<0\), whereas in mean-field the system stays in the normal state (NS).
}
\label{fig:Phases}
\end{figure}

The solutions of the linearized gap equation take the form \cite{SM}:
\[\Delta^{(\ell_x\ell_y\nu)}(\mathbf{p})=\sum_m D^{(\ell_x\ell_y\nu)}_{m}\Phi^{(\ell_x\ell_y\nu)}_m(\mathbf{p})\label{DeltaForm}\]
where \(D^{(\ell_x\ell_y\nu)}_{m}\) are coefficients satisfying
\[D^{(\ell_x\ell_y\nu)}_{m}=-\sum_{m'}g^{(\ell_x\ell_y\nu)}_{m'}\tilde{\Pi}^{(\ell_x\ell_y\nu)}_{mm'}D^{(\ell_x\ell_y\nu)}_{m'}\,.\label{RedLinGapEq-main}\]
This is a matrix equation with
\(\tilde{\Pi}^{(\ell_x\ell_y\nu)}_{mm'}=\sum_\mathbf{k}\Pi(\mathbf{k})\Phi^{(\ell_x\ell_y\mu)}_m(\mathbf{k})\Phi^{(\ell_x\ell_y\nu)}_{m'}(\mathbf{k})\).
We solve Eq. (\ref{RedLinGapEq-main}) numerically to obtain the phase diagram shown in Fig. \ref{fig:Phases} (a). The mean-field analysis uncovers three regimes. When both \(U\) and \(V\) are positive, i.e. there are only repulsive interactions, we find that there is no pairing instability, so the system remains a metal.
For negative (attractive) \(U\), the \((00s;s)\) uniform SC channel in the \(A_g\) irrep always has the highest \(T_c\) irrespective of the sign of $V$. 
However, an interesting situation arises for positive \(U\) and negative (attractive) \(V\),
where we identify the triplet PDW channels \((01t)\) and \((10t)\) as the leading instabilities with the same highest \(T_c\). That is, the two solutions are degenerate and form two components of the 2D irrep \(\mathbf{Q}^*_t\) of the space group, being mapped to each other by the \(\hat{C}_4\) symmetry. This means that any linear combination of the solutions \(\Delta^{(01t)}\) and \(\Delta^{(10t)}\) satisfy the linearized gap equation.

\begin{figure}[t]
\centering
\includegraphics[width=0.49\textwidth]{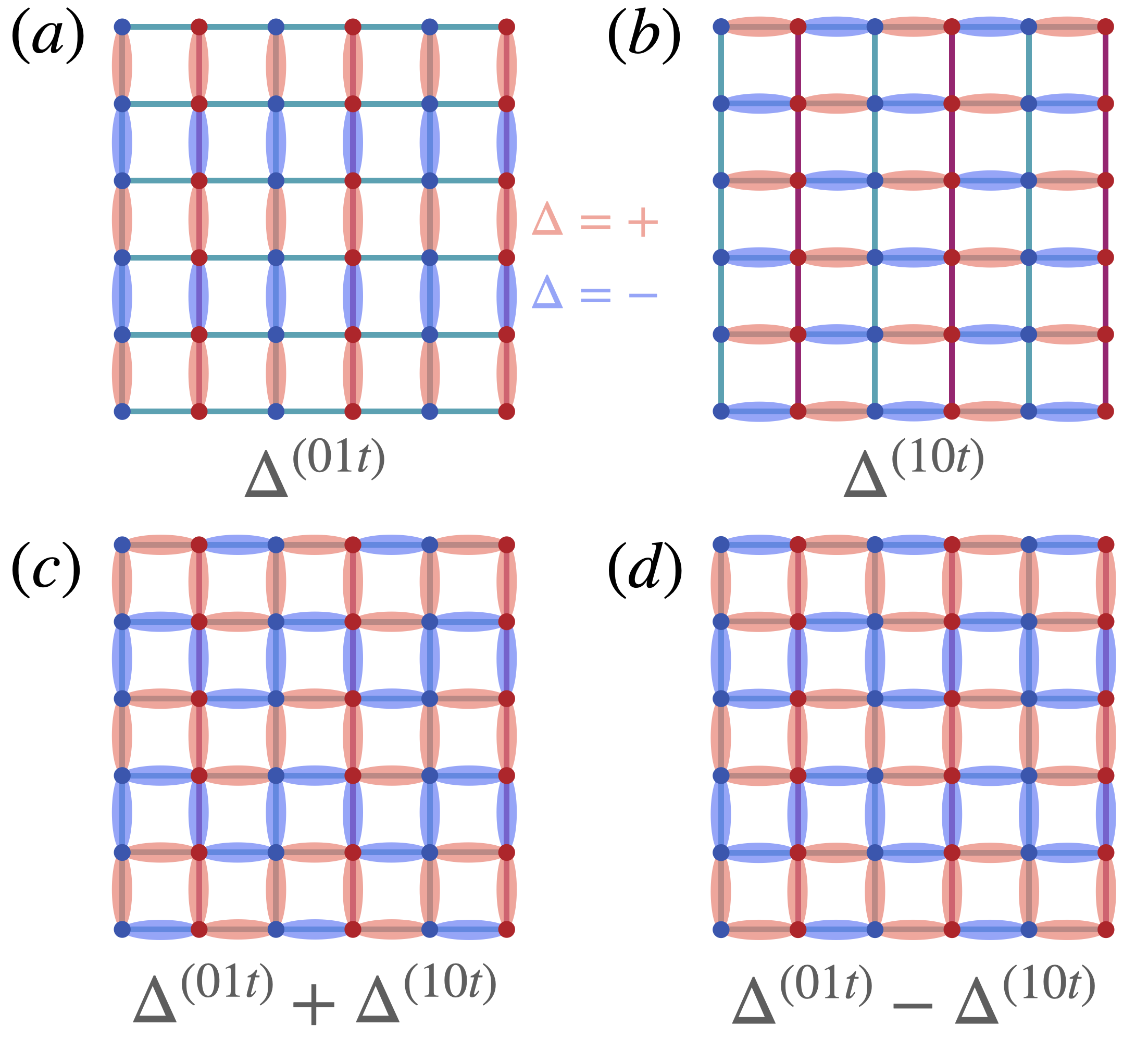}
\caption{Triplet PDW order parameters in real space obtained from minimizing the fourth order free energy Eq. (\ref{F4}). (a) and (b) show the two nodal unidirectional PDW phases, \(\Delta^{(01t)}\) and \(\Delta^{(10t)}\) respectively. (c) and (d) shows the two fully gapped bidirectional PDW phases that are linear combinations of the nodal phases.
}
\label{fig:Orders}
\end{figure}

This degeneracy is lifted by the fourth order terms in the free energy, which in our case is similar to the one considered in \cite{Agterberg08}. Within our model, we compute the fourth order term to be (omitting the \(\nu=t\) index for clarity)
\begin{align}\label{F4}
    \mathcal{F}^{(4)}&=\beta_0(|D^{(01)}|^2+|D^{(10)}|^2)^2+2\beta_1|D^{(01)}|^2|D^{(10)}|^2-\nonumber\\
    &-\beta_1\left((D^{(01)}D^{(10)*})^2+c.c.\right)
\end{align}
with \(\beta_0,\beta_1>0\) \cite{SM}. The free energy is symmetric under \(\hat{C}_4\) that takes \(D^{(01)}\rightarrow D^{(10)}\rightarrow -D^{(01)}\). The minimum of the free energy is degenerate between a unidirectional PDW in which only one of \(D^{(01)}\) or \(D^{(10)}\) is non-zero (breaking the \(\hat{C}_4\) symmetry), and bidirectional combinations with \(D^{(01)}=\pm D^{(10)}\); in both cases TRS is preserved. 
Additional terms neglected in our model may favor either the unidirectional or bidirectional combinations. See Fig. \ref{fig:Orders} that shows the corresponding real-space order parameters. The triplet PDW gap functions have the form \(\Delta^{(01t)}\propto \sin p_y\) and \(\Delta^{(10t)}\propto \sin p_x\), resulting in a nodal fermionic excitation spectrum in the unidirectional phase, but a fully gapped spectrum in the bidirectional case. Note that for the \(\ell_x=1\) unidirectional and the bidirectional phases the MBZ is further folded along the \(p_y\) direction, with \(p_y\in(-\pi/2,\pi/2]\). Though \(\hat{T}_x\) is broken in that case, the ground states still have a \(\mathbb{Z}_2\) symmetry that is a combination of MTG symmetries and a \(U(1)\) transformation, resulting in a two-fold degenerate excitation spectrum.

\textit{Renormalization Group Analysis --}
A limitation of the self-consistent mean-field analysis is that it does not consider all possible instabilities of the system, including possible charge (CDW) and spin density wave (SDW) instabilities. To confirm that the phases found above are true ground states of the system we therefore carry out an additional patch RG analysis following \cite{ShafferWangSantos22} that extended the standard VHS patch RG framework used in the context of cuprates \cite{schulz1987superconductivity,dzyaloshinskiui1987maximal} and doped graphene \cite{Nandkishore2012} to the Hofstadter-Hubbard model with on-site interactions. Here we 
use the same framework extended to include nearest neighbor interactions \cite{SM}.

In this framework we consider the values of the gap functions and the interactions only in the vicinity of the saddle points that give rise to the VHS (see Fig. \ref{fig:Model}), since the density of states diverges at these points. As a result the patch RG has a limitation in juxtaposition to the mean-field analysis as it only determines the gap function at the VHS points. As one consequence, we find that in RG the triplet PDW channel is further degenerate with the \((11s)\) singlet PDW that is odd with respect to both \(\hat{T}_x\) and \(\hat{T}_y\), and further transforms as a 2D irrep \(E_g\) of the point group. However, the mean field analysis shows that this degeneracy is lifted already at the level of the linearized gap equation in favor of the triplet PDW channel. The two methods therefore complement each other.

The resulting RG phase diagram is shown in Fig. \ref{fig:Phases} (b). The main difference from the mean-field phase diagram is the appearance of a \(d\)-wave \((00s;d)\) uniform SC channel belonging to the \(B_g\) irrep when all interactions are repulsive, consistent with the results in \cite{ShafferWangSantos22}. The \(d\)-wave SC moreover competes with the triplet PDW phase for negative \(V\), but the PDW remains the winning instability as long as \(0.21U<-V\), within numerical accuracy. Importantly, the RG analysis rules out the CDW and SDW instabilities and establishes that the triplet PDW is indeed the leading instability in a large portion of the phase diagram.

\textit{Discussion --} In this work we showed that a symmetry-protected triplet PDW is realized at weak-coupling in the \(\pi\)-flux lattice with repulsive on-site interactions and moderate nearest-neighbor attraction. The symmetries that protect this PDW are the magnetic translation symmetries that are characteristic of Hofstadter systems, including the TR-symmetric Hofstadter model \cite{CocksHofstetter12}, of which the TR-symmetric \(\pi\)-flux model is a special case. The experimental realization of a host of Hofstadter bands in moiré \cite{Dean13,Ponomarenko13,Hunt13, Saito21, YuKivelsonFeldman22}, as well as proposed realization using synthetic gauge fields in optical lattices \cite{AidelsburgerBloch13,Cooper19, Lauria22}, open promising routes for realizing PDW states. Furthermore, twisted double-layer copper-oxides may be rich platforms to pursue extensions of this mechanism to the intermediate coupling regime \cite{yu2019high,Can2021}. 

Moreover, this finding opens interesting new research directions to pursue other phenomena that emerge from PDW states. For instance, it will be important to investigate the nature of induced, vestigial, and intertwined orders associated with this new triplet PDW phase, and to draw comparisons with the class of PDW states introduced as a way to explain cuprate phase diagrams
\cite{FradkinKivelsonTranquada15,Agterberg20}. These can interact in non-trivial ways with PDW vortices which may be fractional \cite{Agterberg08} or give rise to exotic charge-4e condensates \cite{BergFradkinKivelson09}.  Moreover, extensions of the present model may uncover mechanisms that lift the degeneracy between the gapless unidirectional and gapped bidirectional PDWs, which may include disorder or effects associated with the gapless fermionic excitations that may be gapped out by additional interactions \cite{SantosWangFradkin19}.
We leave these open questions for future studies.

\section*{Acknowledgments}
We thank Jian Wang for useful discussions and earlier collaborations. This research is supported by the U.S. Department of Energy, Office of Science, Basic Energy Sciences, under Award DE-SC0023327
and by startup funds at Emory University.

\bibliographystyle{apsrev4-1}
\bibliography{bibliography}

\pagebreak
\widetext
\pagebreak
\begin{center}
\textbf{\large Supplementary Material for Triplet Pair-density Wave Superconductivty on the $\pi$-flux Square Lattice}
\end{center}
\setcounter{equation}{0}
\setcounter{figure}{0}
\setcounter{table}{0}
\setcounter{page}{1}
\makeatletter
\renewcommand{\theequation}{S\arabic{equation}}
\renewcommand{\thefigure}{S\arabic{figure}}
\renewcommand{\bibnumfmt}[1]{[S#1]}
\renewcommand{\citenumfont}[1]{S#1}

\section{Interaction Projections}

Here we go over some details of the projection of the interactions in Eq. (\ref{HI}). In momentum space in the sublattice basis
\[c_{\mathbf{k}s\sigma}=\frac{1}{\sqrt{N}}\sum_{\mathbf{R}}e^{-i\mathbf{k}\cdot (s\hat{\mathbf{x}}+\mathbf{R})}c_{s\hat{\mathbf{x}}+\mathbf{R},\sigma}\label{cFT}\]
the on-site interaction Hamiltonian is
\[H_U=U\sum_{\mathbf{k,p,q}s\sigma}c^\dagger_{\mathbf{p+q},s\sigma}c^\dagger_{-\mathbf{p}s,-\sigma}c_{-\mathbf{k}s,-\sigma}c_{\mathbf{k+q},s\sigma}\, ,\]
while the nearest-neighbor interaction Hamiltonian is
\begin{align} 
H_V&=2V\sum_{\mathbf{k,p,q}s\sigma\sigma'}\bigg[\cos(p_y-k_y)c^\dagger_{\mathbf{p+q},s\sigma}c^\dagger_{-\mathbf{p}s\sigma'}c_{-\mathbf{k}s\sigma'}c_{\mathbf{k+q},s\sigma}+\cos(p_x-k_x)c^\dagger_{\mathbf{p+q},s\sigma}c^\dagger_{-\mathbf{p},s+1,\sigma'}c_{-\mathbf{k},s+1,\sigma'}c_{\mathbf{k+q},s\sigma}\bigg]
\end{align}
In the basis Eq. (\ref{d}), the operators \(c_{\mathbf{k}s}\) are
\[c_{\mathbf{k}s}=\frac{1}{\sqrt2}\left[(-1)^s\sqrt{1-(-1)^s\frac{\cos k_y}{E(\mathbf{k})}}d_{\mathbf{k}+}+\sqrt{1+(-1)^s\frac{\cos k_y}{E(\mathbf{k})}}d_{\mathbf{k}-}\right]\label{cInv}\]
To project the interactions, we insert Eq. (\ref{cInv}) into the interaction Hamiltonian and restrict \(\alpha\) to a single value for the desired band, i.e.
\[c_{\mathbf{k}s}\rightarrow\frac{(-\alpha)^s}{\sqrt2}\sqrt{1-\alpha(-1)^s\frac{\cos k_y}{E(\mathbf{k})}}d_{\mathbf{k}\alpha}\label{cProj}\]
and sum over \(s\).

We are interested in the \(\mathbf{q}=0\) and \(\mathbf{q}=\mathbf{Q}=\pi\hat{\mathbf{y}}\) cases corresponding to the \(\ell_x=0,1\) pairing channels, which in the band-basis are
\[H_{int,\alpha}^{(\ell)}=\sum_{\mathbf{k p}\sigma\sigma'} g^{(\ell;\alpha)}(\mathbf{p;k})d^\dagger_{\mathbf{p+\ell\mathbf{Q}},\alpha\sigma}d^\dagger_{-\mathbf{p}\alpha\sigma'}d_{-\mathbf{k}\alpha\sigma'}d_{\mathbf{k+\ell Q},\alpha\sigma}\]
The relevant projections for \(\ell_x=0\) are
\begin{align}
\sum_s c^\dagger_{\mathbf{p}s\sigma}c^\dagger_{-\mathbf{p}s\sigma'}c_{-\mathbf{k}s\sigma'}c_{\mathbf{k}s\sigma}&\rightarrow\frac{1}{4}\sum_s\left(1-\alpha(-1)^s\frac{\cos p_y}{E(\mathbf{p})}\right)\left(1-\alpha(-1)^s\frac{\cos k_y}{E(\mathbf{k})}\right) d^\dagger_{\mathbf{p}\alpha\sigma}d^\dagger_{-\mathbf{p}\alpha\sigma'}d_{-\mathbf{k}\alpha\sigma'}d_{\mathbf{k}\alpha\sigma}=\nonumber\\
&=\frac{1}{2}\left(1+\frac{\cos p_y}{E(\mathbf{p})}\frac{\cos k_y}{E(\mathbf{k})}\right) d^\dagger_{\mathbf{p}\alpha\sigma}d^\dagger_{-\mathbf{p}\alpha\sigma'}d_{-\mathbf{k}\alpha\sigma'}d_{\mathbf{k}\alpha\sigma}
\end{align}
\begin{align}
\sum_s c^\dagger_{\mathbf{p}s\sigma}c^\dagger_{-\mathbf{p},s+1,\sigma'}c_{-\mathbf{k},s+1,\sigma'}c_{\mathbf{k}s\sigma}&\rightarrow\frac{1}{4}\sum_s\sqrt{1-\frac{\cos^2 p_y}{E^2(\mathbf{p})}}\sqrt{1-\frac{\cos^2 k_y}{E^2(\mathbf{k})}} d^\dagger_{\mathbf{p}\alpha\sigma}d^\dagger_{-\mathbf{p}\alpha\sigma'}d_{-\mathbf{k}\alpha\sigma'}d_{\mathbf{k}\alpha\sigma}=\nonumber\\
&=\frac{1}{2}\frac{\cos p_x}{E(\mathbf{p})}\frac{\cos k_x}{E(\mathbf{k})} d^\dagger_{\mathbf{p}\alpha\sigma}d^\dagger_{-\mathbf{p}\alpha\sigma'}d_{-\mathbf{k}\alpha\sigma'}d_{\mathbf{k}\alpha\sigma}
\end{align}
For \(\ell_x=1\), the projections are
\begin{align}
\sum_s c^\dagger_{\mathbf{p+Q},s\sigma}c^\dagger_{-\mathbf{p}s\sigma'}c_{-\mathbf{k}s\sigma'}c_{\mathbf{k+Q},s\sigma}&\rightarrow\frac{1}{4}\sum_s\sqrt{1-\frac{\cos^2 p_y}{E^2(\mathbf{p})}}\sqrt{1-\frac{\cos^2 k_y}{E^2(\mathbf{k})}} d^\dagger_{\mathbf{p}\alpha\sigma}d^\dagger_{-\mathbf{p}\alpha\sigma'}d_{-\mathbf{k}\alpha\sigma'}d_{\mathbf{k}\alpha\sigma}=\nonumber\\
&=\frac{1}{2}\frac{|\cos p_x|}{E(\mathbf{p})}\frac{|\cos k_x|}{E(\mathbf{k})} d^\dagger_{\mathbf{p}\alpha\sigma}d^\dagger_{-\mathbf{p}\alpha\sigma'}d_{-\mathbf{k}\alpha\sigma'}d_{\mathbf{k}\alpha\sigma}\\
\sum_s c^\dagger_{\mathbf{p+Q},s\sigma}c^\dagger_{-\mathbf{p},s+1,\sigma'}c_{-\mathbf{k},s+1,\sigma'}c_{\mathbf{k+Q},s\sigma}&\rightarrow\frac{1}{4}\sum_s
\left(1+\alpha(-1)^s\frac{\cos p_y}{E(\mathbf{p})}\right)\left(1+\alpha(-1)^s\frac{\cos k_y}{E(\mathbf{k})}\right)d^\dagger_{\mathbf{p}\alpha\sigma}d^\dagger_{-\mathbf{p}\alpha\sigma'}d_{-\mathbf{k}\alpha\sigma'}d_{\mathbf{k}\alpha\sigma}=\nonumber\\
&=\frac{1}{2}
\left(1+\frac{\cos p_y}{E(\mathbf{p})}\frac{\cos k_y}{E(\mathbf{k})}\right) d^\dagger_{\mathbf{p}\alpha\sigma}d^\dagger_{-\mathbf{p}\alpha\sigma'}d_{-\mathbf{k}\alpha\sigma'}d_{\mathbf{k}\alpha\sigma}
\end{align}
Note that none of the form factors depend on the band index \(\alpha\), which we therefore drop from now on.

Next, we switch from the charge/spin decomposition of the interactions to the singlet/triplet pairing decomposition, using the Pauli matrix completeness relation
\[2\delta_{\sigma_1\sigma_2'}\delta_{\sigma'_1\sigma_2}=\sum_\nu \sigma^\nu_{\sigma_1\sigma_1'}\sigma^\nu_{\sigma_2\sigma_2'}=\sum_\nu (\sigma^\nu i\sigma^y)^*_{\sigma_1\sigma_1'}(\sigma^\nu i\sigma^y)_{\sigma_2\sigma_2'}\]
where \(\nu=0,x,y,z\) with \(0\) corresponding to singlet and the rest to triplet pairing channels; this puts the projected interactions into the following form:
\[H^{(\ell_x\nu)}_{int}=\frac{1}{2}\sum_{\mathbf{k p}\sigma_1\sigma_1'\sigma_2\sigma_2'} g^{(\ell_x\nu)}(\mathbf{p;k})(\sigma^\nu i\sigma^y)^*_{\sigma_1\sigma_1'}(\sigma^\nu i\sigma^y)_{\sigma_2\sigma_2'}d^\dagger_{\mathbf{p}+\ell_x\mathbf{Q},\alpha\sigma_1}d^\dagger_{-\mathbf{p}\alpha\sigma_1'}d_{-\mathbf{k}\alpha\sigma_2}d_{\mathbf{k}+\ell_x\mathbf{ Q},\alpha\sigma'_2}\]
The anti-commutation relations then imply
\[g^{(\ell_x\nu)}(\mathbf{p;k})=(-1)^\nu g^{(\ell_x\nu)}(-\mathbf{p}-\ell_x \mathbf{Q;k})=g^{(\ell_x\nu)}(-\mathbf{p}-\ell_x \mathbf{Q;-k-\ell_x Q})\]
(with \((-1)^{\nu}=-1\) for \(\nu=x,y,z\)), which implies that \(g^{(\ell_x\nu)}(\mathbf{p;k})\) is even or odd under \(\mathbf{p}\rightarrow-\mathbf{p}+\ell_x \mathbf{Q}\) for singlet and triplet channels respectively.

We further split the projected interactions into \(g^{(\ell_x\ell_y\nu)}(\mathbf{p;k})=(-1)^{\ell_y} g^{(\ell_x\ell_y\nu)}(\mathbf{p+Q;k})\) according to channels even and odd under \(\hat{T}_x\), which we know from the irrep analysis must decouple. Finally, the \((00s)\) channel splits into a \(\hat{C}_4\) even and odd parts (i.e. `s'- and `d'-wave channels) that we refer to as \((00s;s)\) and \((00s;d)\). The \(\hat{C}_4\) symmetry acts as
\[\hat{C}_4 c_{\mathbf{p}+\ell_x\mathbf{Q},s\sigma}\hat{C}_4^\dagger=\frac{1}{2}\sum_{s'\ell'_x}(-1)^{ss'+\ell_x s'+\ell'_x s} c_{\bar{\mathbf{p}} +\ell'_x \mathbf{Q},s' \sigma}\,. \label{C4}\]
Using the fact that \(\cos (p_j-k_j)=\cos p_j \cos k_j+\sin p_j \sin k_j\), we find that the projected interactions for each channel are:
\begin{align}
    g^{(00s;s)}(\mathbf{p;k})&=\frac{U}{2}+V\frac{(\cos^2p_y+\cos^2p_x)(\cos^2k_y+\cos^2k_x)}{E(\mathbf{p})E(\mathbf{k})}\nonumber\\
    g^{(00s;d)}(\mathbf{p;k})&=V\frac{(\cos^2p_y-\cos^2p_x)(\cos^2k_y-\cos^2k_x)}{E(\mathbf{p})E(\mathbf{k})}\nonumber\\
    g^{(01s)}(\mathbf{p;k})&=\frac{U}{2}\frac{\cos p_y\cos k_y}{E(\mathbf{p})E(\mathbf{k})}+V\cos p_y\cos k_y\nonumber\\
    g^{(10s)}(\mathbf{p;k})&=\frac{U}{2}\frac{\cos p_x\cos k_x}{E(\mathbf{p})E(\mathbf{k})}+V\cos p_x\cos k_x\nonumber\\
    g^{(11s)}(\mathbf{p;k})&=V\frac{\sin p_y\cos p_x\sin k_y\cos k_x+\sin p_x\cos p_y\sin k_x\cos k_y}{E(\mathbf{p})E(\mathbf{k})}\nonumber\\
    g^{(00t)}(\mathbf{p;k})&=V\frac{\sin p_y\cos p_y\sin k_y\cos k_y+\sin p_x\cos p_x\sin k_x\cos k_x}{E(\mathbf{p})E(\mathbf{k})}\nonumber\\
    g^{(01t)}(\mathbf{p;k})&=V\sin p_y\sin k_y\nonumber\\
    g^{(10t)}(\mathbf{p;k})&=V\sin p_x\sin k_x\nonumber\\
    g^{(11t)}(\mathbf{p;k})&=V\frac{\cos p_x\cos p_y\cos k_x\cos k_y}{E(\mathbf{p})E(\mathbf{k})}
\end{align}
From these we obtain Table \ref{table:irreps}. Observe that \(\pi\hat{\mathbf{y}}\) and \(\pi\hat{\mathbf{x}}\) are mapped to each other by \(\hat{C}_4\) (since it exchanged \(\hat{T}_x\) and \(\hat{T}_y\)), which implies that \((01\nu)\) and \((10\nu)\) channels form two components of the same space group irrep; the little point group is trivial in this case since \(\hat{C}_4\) symmetry is broken. The irreps of \((00\nu)\) and \((11\nu)\), on the other hand, have a trivial star but a non-trivial little point group, and are therefore classified according to the irreps of \(D_{4h}\) (since inversion symmetry is also present). In particular, the \((00s;s)\) channel corresponds to the trivial irrep \(A_{1g}\), \((00s;d)\) to the \(B_{1g}\) irrep, \((11s)\) to the \(E_{g}\) irrep, \((00t)\) to the \(E_{u}\) irrep, and \((11t)\) to the \(A_{1u}\) irrep.

An astute reader may notice that the \(g^{(10\nu)}\) and \(g^{(11\nu)}\) are discontinuous on the MBZ due to factors of \(\cos k_x\), which evaluates to \(\pm 1\) for \(k_x=\pm \pi/2\) respectively on the MBZ boundary. However, this is only an apparent discontinuity that is due to the discontinuity of the \(d_{\mathbf{k}\alpha}\), a consequence of the presence of Dirac nodes in the system that imply an existence of a branch cut in the wavefunction. Note in particular that the operator \(c_{\mathbf{k}1}\) has a branch cut at the \(k_x=\pm \pi/2\) because, as follows from the definition in Eq. (\ref{cFT})
\[c_{\mathbf{k}+\pi\hat{\mathbf{x}},s}=e^{-ik_x s}c_{\mathbf{k}s}\]
so that technically \(c_{(\pi/2,k_y),1}=-c_{(-\pi/2,k_y),1}\). For that reason we need to consider \(k_x=\pi/2\) and \(k_x=-\pi/2\) as distinct points. The operators \(d_{\mathbf{k}\alpha}\) can be made continuous for \(\alpha\cos k_y<0\) if we take
\[d_{\mathbf{k}\alpha}=\frac{1}{\sqrt2}\sqrt{1+\alpha\frac{\cos k_y}{E(\mathbf{k})}}c_{\mathbf{k}0}+\frac{\alpha\,\sgn[\cos k_x]}{\sqrt2}\sqrt{1-\alpha\frac{\cos k_y}{E(\mathbf{k})}}c_{\mathbf{k}1}\label{dsgn}\]
so that \(d_{\mathbf{k}+\pi\hat{\mathbf{x}},\alpha}=d_{\mathbf{k}\alpha}\) for \(\alpha\cos k_y<0\). For \(\alpha\cos k_y>0\), however, we then get \(d_{(\pi/2,k_y)\alpha} = \alpha\,\sgn[\cos(\pi/2)]c_{(\pi/2,k_y)1}= -\alpha\,\sgn[\cos(\pi/2)]c_{(-\pi/2,k_y)1}= -\frac{\sgn[\cos(\pi/2)]}{\sgn[\cos(-\pi/2)]}d_{(-\pi/2,k_y)\alpha}\). To be consistent, one is forced to take \(d_{(\pi/2,k_y)\alpha}=-d_{(-\pi/2,k_y)\alpha}\) for \(\alpha\cos k_y>0\), with the coefficients in Eq. (\ref{dsgn}) being continuous.

Taking this into account, we note that while anti-commutation relations typically imply that triplet interaction must vanish when \(\mathbf{p}=-\mathbf{p+\ell_x Q}\), one has to be careful in the case \(p_x=\pm \pi/2\) because of the fact that \(d_{(\pi/2,k_y)\alpha}=-d_{(-\pi/2,k_y)\alpha}\) for \(\alpha\cos k_y>0\). This makes no difference for \(\ell_x=0\) since the minus signs cancel, but for \(\ell_x=1\) there is an additional minus sign that means that
\[g^{(1\nu)}((\pi/2,k_y);\mathbf{k})=-(-1)^\nu g^{(1\nu)}((\pi/2,k_y);\mathbf{k})\]
i.e. it is the singlet projected interactions that must vanish while the triplet interaction does not, consistent with our results.

\section{Some Details of the Gap Equation}

Here we provide a few details about solving the gap equation Eq. (\ref{RedLinGapEq-main}). In most cases, there is a single component \(m=0\), and we simply evaluate the integral on the RHS. The only non-trivial cases are for the \((00s;s)\), \((01s)\) and \((10s)\) channels, for which the reduced gap equation becomes a \(2\times2\) matrix equation admitting two solutions in each channel. In particular, for the \((00s;s)\) channel we have
\[\left(\begin{array}{c}
    D^{(00s;s)}_{0}  \\
    D^{(00s;s)}_{1} 
\end{array}\right)=-\left(\begin{array}{cc}
    U \tilde{\Pi}^{(00s;s)}_{00}/2 & U \tilde{\Pi}^{(00s;s)}_{01}/2  \\
    V \tilde{\Pi}^{(00s;s)}_{01}  & V \tilde{\Pi}^{(00s;s)}_{11}
\end{array}\right)\left(\begin{array}{c}
    D^{(00s;s)}_{0}  \\
    D^{(00s;s)}_{1} 
\end{array}\right)\]
The matrix has eigenvalues
\[\gamma^{(00s;s)}_\pm=\frac{1}{4}\left[U \tilde{\Pi}^{(00s;s)}_{00}+2V \tilde{\Pi}^{(00s;s)}_{11}\mp\sqrt{\left(U \tilde{\Pi}^{(00s;s)}_{00}-2V \tilde{\Pi}^{(00s;s)}_{11}\right)^2+8 U V\left(\tilde{\Pi}^{(00s;s)}_{01}\right)^2}\right]\]
with eigenstates that determine the relative weights of \(D^{(00s;s)}_{0}\) and \(D^{(00s;s)}_{1}\):
\[\left(\begin{array}{c}
    D^{(00s;s)}_{0,\pm}  \\
    D^{(00s;s)}_{1,\pm} 
\end{array}\right)\propto\left(\begin{array}{c}
    U \tilde{\Pi}^{(00s;s)}_{00}+2V \tilde{\Pi}^{(00s;s)}_{11}\mp\sqrt{\left(U \tilde{\Pi}^{(00s;s)}_{00}-2V \tilde{\Pi}^{(00s;s)}_{11}\right)^2+8 U V\left(\tilde{\Pi}^{(00s;s)}_{01}\right)^2}  \\
    4 V \tilde{\Pi}^{(00s;s)}_{01}
\end{array}\right) \]
(note that the linearized gap equation does not fix the magnitude of the \(D^{(\ell_x\ell_y\nu)}_{m}\) coefficients). The equations are similar for the \((01s)\) and \((10s)\) channels, which are degenerate as they belong to the same space group irrep.

In the end we thus obtain the self-consistency relations
\[1=-\gamma^{(\ell_x\ell_y\nu)}(T)\]
which is an equation for \(T_c\) since the RHS is a function of the temperature \(T\). At the phase transition, the instability happens only for the channels with the highest \(T_c\), and we can ignore the rest. In particular, since \(\gamma^{(00s;s)}_+<\gamma^{(00s;s)}_-\), the \(\gamma^{(00s;s)}_-\) channel always gives a lower \(T_c\) and it can be ignored; we thus only keep the + solution and can drop the \(\pm\) index. To obtain the mean-field phase diagram, we select the channel with largest eigenvalue (numerically we observe that which eigenvalue is largest appears to be independent of the temperature).

\section{Fourth Order Free Energy}

When the winning irrep is 2D, in order to determine the symmetries of the ground state it is necessary to go to the fourth order in the free energy. The Ginzburg-Landau free energy for Hofstadter superconductors with arbitrary flux has been derived in \cite{ShafferWangSantos21}, here we summarize the simplified case for \(\pi\)-flux. Defining the matrix gap function \(\hat{\Delta}(\mathbf{p})\) via
\[H_{SC}=\sum\hat{\Delta}_{\ell\ell';\sigma\sigma'}(\mathbf{p})d^\dagger_{\mathbf{p},\ell,\sigma}d^\dagger_{\mathbf{p},\ell,\sigma'}\]
with \(\mathbf{p}\) restricted to the reduced MBZ, the fourth order term in the free energy can be computed as
\[\mathcal{F}^{(4)}=\sum_\mathbf{k}\beta(\mathbf{p})\text{Tr}\left[\hat{\Delta}^\dagger(\mathbf{p})\hat{\Delta}(\mathbf{p})\hat{\Delta}^\dagger(\mathbf{p})\hat{\Delta}(\mathbf{p})\right]\]
with the trace taken over both the magnetic patch indices and the spin indices, and where
\[\beta(\mathbf{p})=\sum_n \frac{1}{(\omega_n^2+\epsilon^2(\mathbf{p}))^2}=\frac{T\tanh\left(\frac{\epsilon}{2T}\right)-\epsilon\sech^2\left(\frac{\epsilon}{2T}\right)}{8\epsilon^3T^2}\]
Note that in the case of interest (for the \(\mathbf{Q}^*_t\) triplet PDW irrep)
\[\hat{\Delta}=\left(\begin{array}{cc}
    \Delta^{(01)}(\mathbf{p}) & \Delta^{(10)}(\mathbf{p}) \\
    \Delta^{(10)}(\mathbf{p}) & -\Delta^{(01)}(\mathbf{p})
\end{array}\right)\]
and the trace evaluates to
\[\text{Tr}\left[\hat{\Delta}^\dagger(\mathbf{p})\hat{\Delta}(\mathbf{p})\hat{\Delta}^\dagger(\mathbf{p})\hat{\Delta}(\mathbf{p})\right]=\frac{1}{4}\left[(|\Delta^{(01)}|^2+|\Delta^{(10)}|^2)^2+2|\Delta^{(01)}|^2|\Delta^{(10)}|^2-\left((\Delta^{(01)})^2(\Delta^{(10)*})^2+c.c.\right)\right]\]
After the sum over momentum, the free energy is written in terms of the order parameters \(D^{(\ell,\pm)}\) in Eq. (\ref{DeltaForm}) as
\[\mathcal{F}^{(4)}=\beta_0(|D^{(01)}|^2+|D^{(10)}|^2)^2+2\beta_1|D^{(01)}|^2|D^{(10)}|^2-\beta_1\left((D^{(01)})^2(D^{(10)*})^2+c.c.\right)\]
The minimum of the free energy is degenerate between unidirectional combinations in which only one of \(D^{(01)}\) or \(D^{(10)}\) is non-zero (breaking TRS and \(\hat{C}_4\) symmetry), and bidirectional combinations where \(D^{(01)}=\pm D^{(10)}\). Additional terms neglected in our model may favor unidirectional or bidirectional combinations.

\section{Details of RG Calculation}

\begin{figure}[h!]
\centering
\includegraphics[width=0.75 \textwidth]{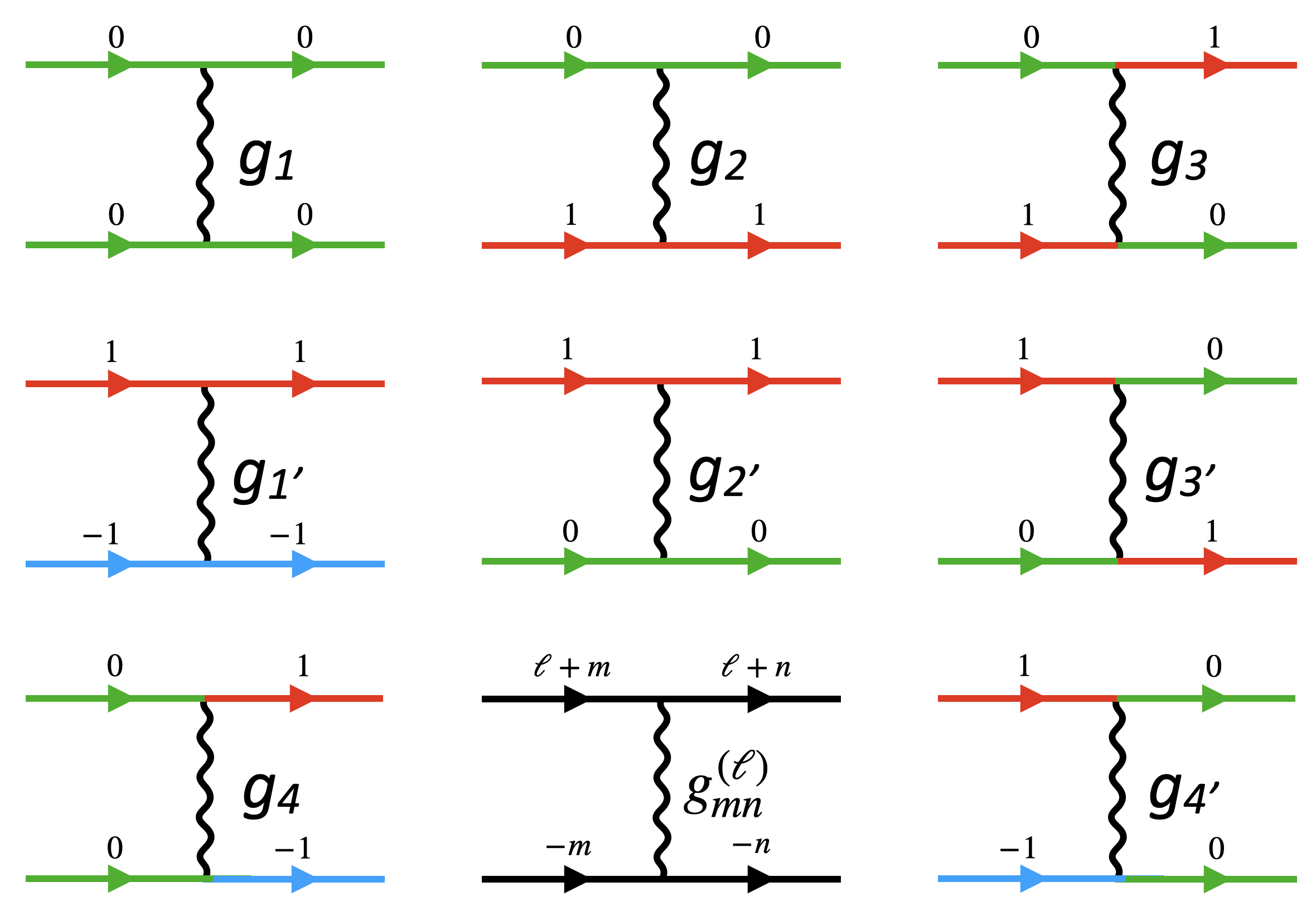}
\caption{Feynman diagrams corresponding to interactions in Eq. (\ref{HintVHS}). The colors indicate the VHS indices for the intrapatch processes \(g_j\) (green lines correspond to \(\mathrm{v}=0\), red to \(\mathrm{v}=1\) and blue to \(\mathrm{v}=-1\)). The colorless diagram shows the interpatch indices with \(\ell\) labeling the total momentum of the incoming and outgoing pairs, and \(m\) and \(n\) labeling the relative momenta of incoming and outgoing pairs respectively.} 
\label{fig:g}
\end{figure}

We use the RG flow equations we derived in \cite{ShafferWangSantos22} (see supplementary material of that reference), which we reproduce here for the special case of the \(\pi\)-flux lattice. In the patch RG formalism we consider states close to the VHS points \(\mathbf{K}_{\ell_x,\mathrm{v}}=\left((1+\mathrm{v})\frac{\pi}{2},\mathrm{v}\frac{\pi}{2}\right)+\ell_x\mathbf{Q}\) where we introduce the VHS index \(\mathrm{v}=0,1\) (for simplicity we will set \(\ell_x=\ell\) defined modulo 2 and we will omit the subscript). Corresponding to these states we define operators \(d_{\mathbf{p}\ell \mathrm{v}\alpha\sigma} = d_{\mathbf{p}+\mathbf{K}_{\ell,\mathrm{v}}, \alpha,\sigma}\) with \(\mathbf{p}\) a small momentum expanded around a patch centered at \(\mathbf{K}_{\ell,\mathrm{v}}\). For bookkeeping purposes it will also be convenient to introduce a redundant VHS index \(\mathrm{v}=-1\), with \(\mathbf{K}_{\ell,-1}=\mathbf{K}_{\ell-1,1}\). We similarly consider the interactions in the vicinity of these patches:
\[H_{int}\rightarrow H_{int}=\frac{1}{2}\sum_{\substack{\ell m n\\ \mathrm{u} \mathrm{v} \mathrm{w}, \sigma\sigma'}} g^{(\ell,\mathrm{u})}_{m,\mathrm{v};n,\mathrm{w}} d^\dagger_{\ell+n,\mathrm{u}+\mathrm{w},\alpha,\sigma}d^\dagger_{-n,-\mathrm{w},\alpha,\sigma'}d_{-m,-\mathrm{v},\alpha,\sigma'}d_{\ell+m,\mathrm{u}+\mathrm{v},\alpha,\sigma}\label{HintVHS}\,,\]
where \(\ell,m,n=0,1\) are magnetic flavor indices, \(\mathrm{u},\mathrm{v},\mathrm{w}=0,\pm 1\) are additional VHS indices.

In the pairing channel we have \(g^{(\ell,0)}_{m,\mathrm{v};n,\mathrm{w}}=g^{(\ell;\alpha)}(\mathbf{K}_{m,\mathrm{v}};\mathbf{K}_{n,\mathrm{w}})\), and there are additional \(g^{(\ell,1)}_{m,\mathrm{v};n,\mathrm{w}}\) interactions in the particle-hole channel corresponding to interactions between pairs with total momentum \((\pi/2,\pi/2)+\ell\mathbf{Q}\). We group the interactions according to their VHS indices as follows:
\begin{align}\label{gs}
    g^{(\ell)1}_{mn}&=g^{(\ell,0)}_{m,0;n,0} 
    &g^{(\ell)1'}_{mn}=g^{(\ell,0)}_{m,1;n,1}\nonumber\\
    g^{(\ell)2}_{mn}&=g^{(\ell,1)}_{m,0;n,0}\nonumber
    &g^{(\ell)3}_{mn}=g^{(\ell,1)}_{m,0;n,-1}\nonumber\\
    g^{(\ell)4}_{mn}&=g^{(\ell,0)}_{m,0;n,1}
    \,,
\end{align}
The corresponding Feynman diagrams are shown in Fig. \ref{fig:g}. Note that some of the diagrams are redundant due to hermiticity and commutation relations. In particular,
\begin{align}\label{Grelations}
    g^{(\ell)1}_{mn}&=g^{(\ell),1*}_{nm}=g^{(\ell)1}_{-\ell-m,-\ell-n}\\
    g^{(\ell)1'}_{mn}&=g^{(\ell),1'*}_{nm}=g^{(\ell)1'}_{-1-\ell-m,-1-\ell-n}\nonumber\\
    g^{(\ell)2}_{mn}&=g^{(\ell),2*}_{nm}=g^{(\ell)2'}_{-\ell-m,-\ell-n}\nonumber\\
    g^{(\ell)3}_{mn}&=g^{(\ell),3*}_{-\ell-n,-\ell-m}=g^{(\ell)3'}_{-\ell-m,-\ell-n}\nonumber\\
    g^{(\ell)4}_{mn}&=g^{(\ell),4}_{-\ell-m,-\ell-n-1}=g^{(\ell)4'*}_{nm}\nonumber
\end{align}
These are in addition to the MTG-imposed relation \(g^{(\ell),j}_{mn}=g^{(\ell),j}_{m-1,n-1}\) and TRS that implies that all coupling constants are real, so that in total there are 16 independent coupling constants.

\begin{figure}[b]
\centering
\includegraphics[width=0.75\textwidth]{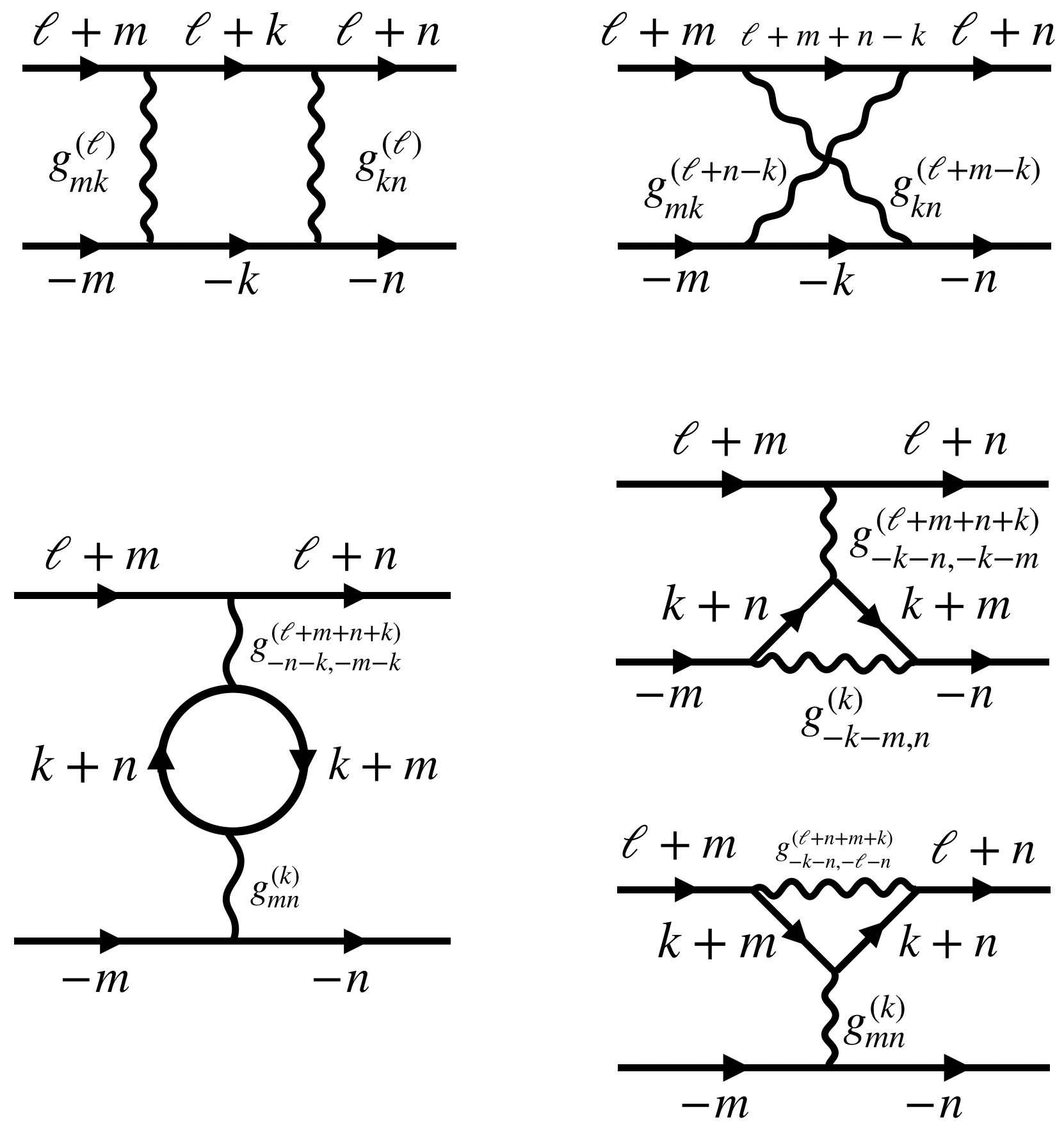}
\caption{Magnetic patch structure of 1 loop RG Feynman diagrams.}
\label{fig:RGflowInt}
\end{figure}

\begin{figure*}[t]
\centering
\includegraphics[width=0.95\textwidth]{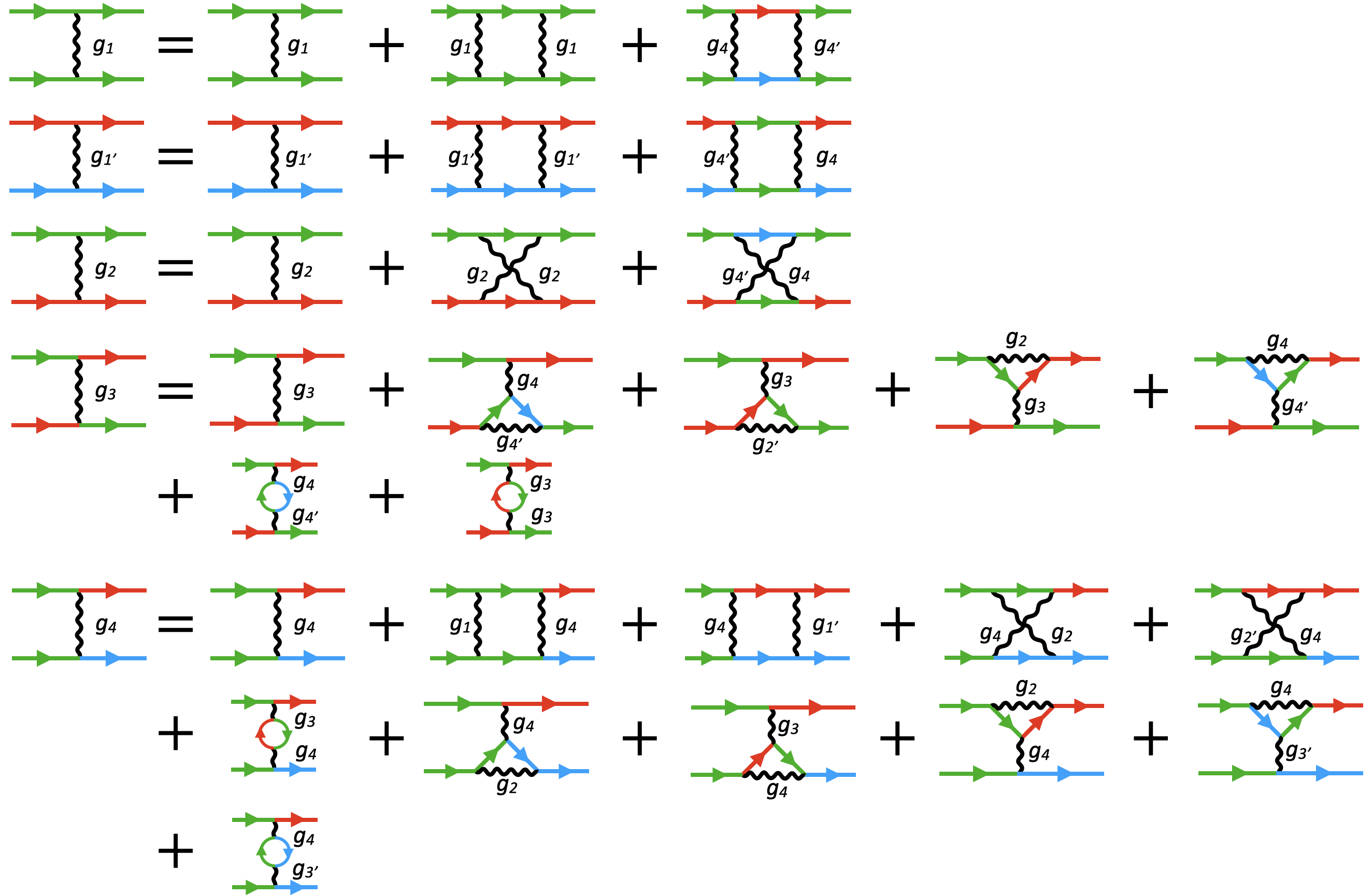}
\caption{The VHS index structure of the 1 loop RG flow diagrams.}
\label{fig:RGflowEq}
\end{figure*}

The basic building blocks of RG are the particle-particle bubble, which also serves as the RG time \(t=\Pi_{pp}^{(0)}\),
\begin{align}
\Pi_{pp}^{(\mathrm{v})}&=-i T\sum_{\omega}\int G_{n,\mathrm{v}}(i\omega,\mathbf{p})G_{\ell-n,\mathrm{v}}(-i\omega,-\mathbf{p})\frac{d^2 p}{(2\pi)^2}=\nu_0 \log^2 \frac{\Lambda}{E}\,,
\end{align}
and the particle-hole bubble is
\begin{align}
\Pi_{ph}^{(\mathrm{v})}&=i T\sum_{\omega}\int G_{n,\mathrm{v}}(i\omega,\mathbf{p})G_{\ell+n,\mathrm{v}+1}(i\omega,\mathbf{p})\frac{d^2 p}{(2\pi)^2}=\nu_0 \log^2 \frac{\Lambda}{E}    
\end{align}
(neither of the bubbles depends on the choice of \(n\) and \(\ell\) by MTG symmetry), where
\[G_{n,\mathrm{v}}(i\omega,\mathbf{p})=\frac{1}{i\omega-\varepsilon_{n,\mathrm{v}}(\mathbf{p})}\]
is the Green's function for the patch around \(\mathbf{K}_{n,\mathrm{v}}\), and \(E\) is the energy scale down to which the high energy modes have been integrated out to. The dispersion expanded around the VHS points is \(\varepsilon_{\ell,\mathrm{v}}(\mathbf{p})\approx \pm (-1)^\mathrm{v}\frac{p_x^2-p_y^2}{2m}-\mu\) (with \(\pm\) for upper and lower bands), and the extra logarithm comes from the diverging DOS at the VHSs. We further define \(d_{pp}^{(\mathrm{v})}=\frac{d\Pi_{pp}^{(\mathrm{v})}}{d\Pi_{pp}^{(0)}}\approx\frac{\Pi_{pp}^{(\mathrm{v})}}{\Pi_{pp}^{(0)}}\) and \(d_{ph}^{(\mathrm{v})}=\frac{d\Pi_{ph}^{(\mathrm{v})}}{d\Pi_{pp}^{(0)}}\approx\frac{\Pi_{ph}^{(\mathrm{v})}}{\Pi_{pp}^{(0)}}\). Note that due to the \(\hat{C}_4\) symmetry \(d_{pp}^{(0)}=d_{pp}^{(1)}=1\) and \(d_{ph}^{(0)}=d_{ph}^{(1)}\). We thus drop the superscripts.

We then obtain the standard 1 loop RG flow equation using the diagrams in Fig. \ref{fig:RGflowEq}, plugging in the magnetic flavor indices from Fig. \ref{fig:RGflowInt}, yielding the RG flow equations:
\begin{align}\label{Gflow}
\dot{g}^{(\ell)1}_{mn}&=-g^{(\ell)1}_{mk}g^{(\ell)1}_{kn}-g^{(\ell)4}_{mk}g^{(\ell)4*}_{nk}\\
\dot{g}^{(\ell)1'}_{mn}&=-g^{(\ell)1'}_{mk}g^{(\ell)1'}_{kn}-g^{(\ell)4*}_{km}g^{(\ell)4}_{kn}\nonumber\\
\dot{g}^{(\ell)2}_{mn}&=d_{ph}\left(g^{(\ell+n-k)2}_{mk}g^{(\ell+m-k)2}_{kn}+g^{(\ell+n-k)4*}_{k,m-1}g^{(\ell+m-k)4}_{k,n-1}\right)\nonumber\\
\dot{g}^{(\ell)3}_{mn}&=2d_{ph}g^{(\ell+m+n+k)3}_{-n-k,-m-k}\left(g^{(k)2}_{m,-n-k}-g^{(k)3}_{mn}\right)+d_{ph}g^{(\ell+m+n+k)4}_{-n-k,-m-k}\left(g^{(k)4*}_{n,-m-k}-2g^{(k)4*}_{n,m-1}\right)+d_{ph}g^{(\ell+m+n+k)4}_{-n-k,-n-\ell-1}g^{(k)4*}_{n,m-1}\nonumber\\
\dot{g}^{(\ell)4}_{mn}&=-g^{(\ell)1}_{mk}g^{(\ell)4}_{kn}-g^{(\ell)4}_{mk}g^{(\ell)1'}_{kn}+d_{ph}\left(g^{(\ell+n-k)2}_{k-\ell-m-n,-\ell-n}g^{(\ell+m-k)4}_{kn}+g^{(\ell+n-k)4}_{mk}g^{(\ell+m-k-1)2}_{k+1,n+1}\right)-\nonumber\\
&+d_{ph}g^{(\ell+m+n+k)4}_{-n-k,-m-k}\left(g^{(k-1)2}_{-m-k+1,n+1}-2g^{(k-1)3}_{1-k-m,-k-n}\right)+d_{ph}g^{(\ell+m+n+k)4}_{-n-k,-n-\ell-1}g^{(k-1)3}_{1-k-m,-k-n}+\nonumber\\
&+d_{ph}g^{(\ell+m+n+k)3}_{-n-k,-m-k}g^{(k)4}_{-m-k,n}+d_{ph}\left(g^{(\ell+m+n+k),2}_{-n-k,-n-\ell}-2g^{(\ell+m+n+k)3}_{-n-k,-m-k}\right)g^{(k)4}_{mn}\nonumber
\end{align}
We then study this flow equation with the bare coupling constants obtained from the projection as the initial condition.

\subsection*{Vertices}

\begin{figure*}[t]
\centering
\includegraphics[width=0.9\textwidth]{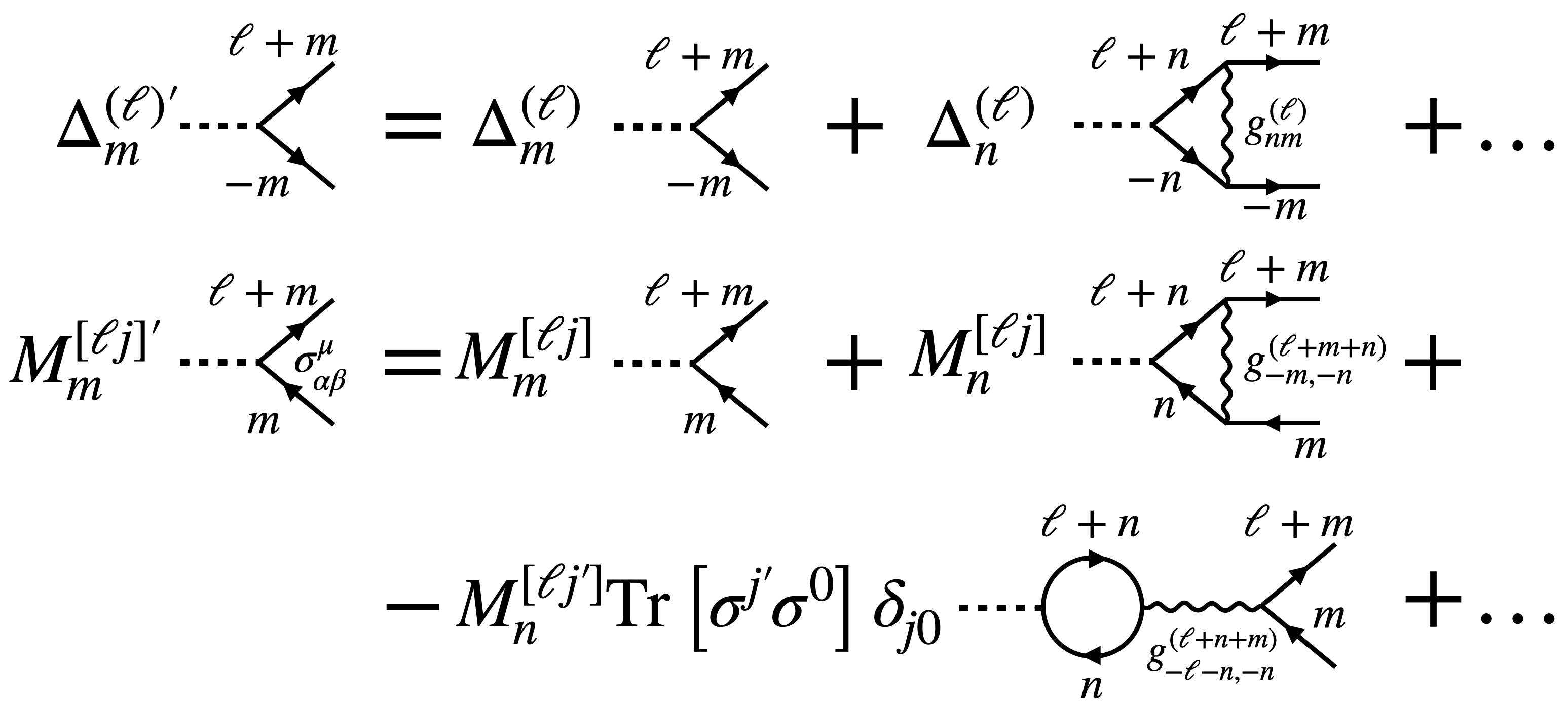}
\caption{Magnetic flavor structure of the 1 loop Feynman diagrams contributing to vertex flow.}
\label{fig:RGflowVertex}
\end{figure*}

\begin{figure*}[t]
\centering
\includegraphics[width=0.95\textwidth]{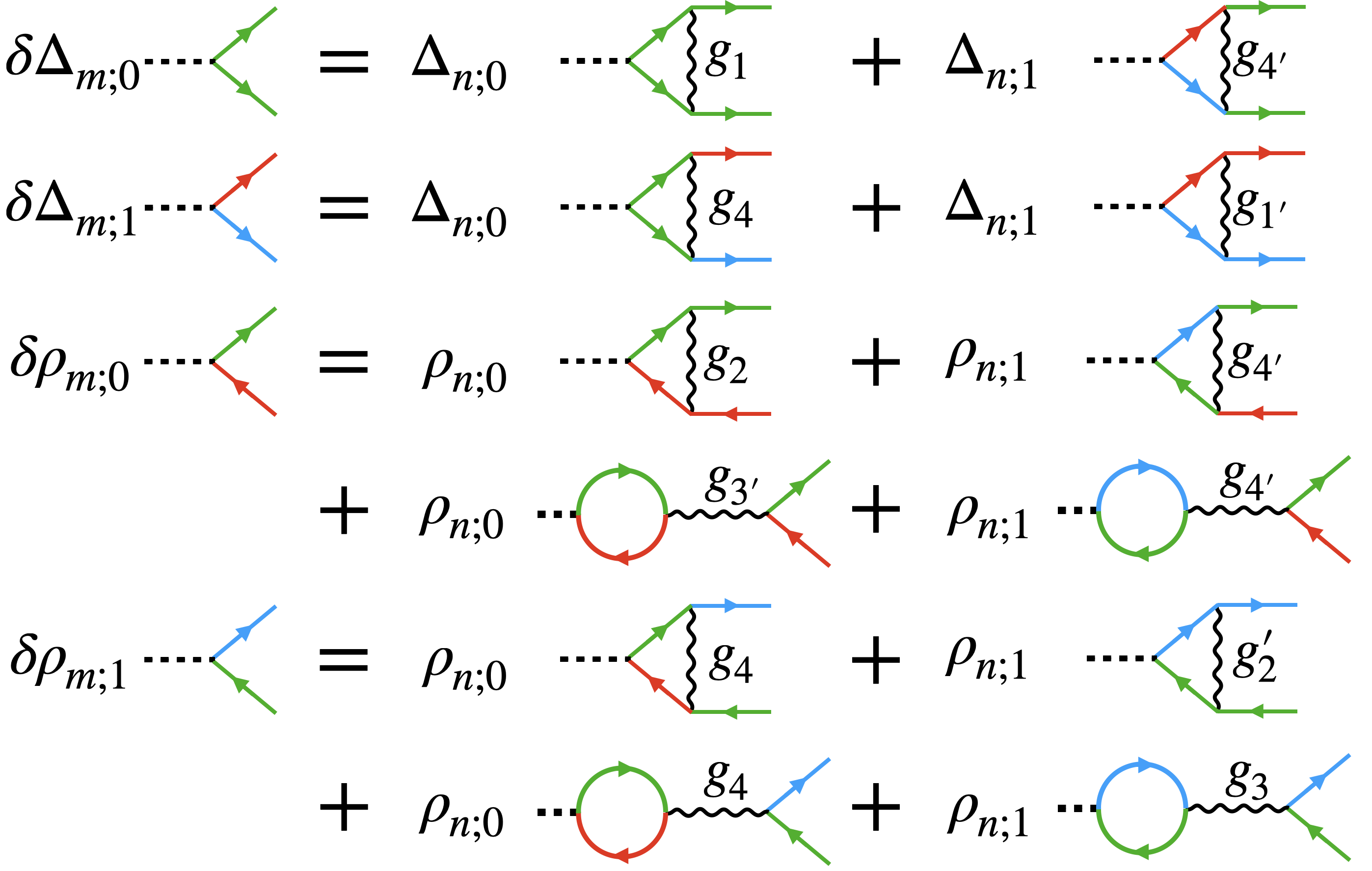}
\caption{VHS index structure of the 1 loop Feynman diagrams contributing to the vertex flow (green for \(\mathrm{v}=0\), red/blue for \(\mathrm{v}=\pm1\) respectively). SDW diagrams are the same as CDW diagrams with \(M\) instead of \(\rho\).}
\label{fig:RGflowVertexEq}
\end{figure*}

In order to determine which instability is realized under the RG flow, we study the flow of the associated susceptibilities and vertices. The relevant vertex Hamiltonians are
\begin{eqnarray}\label{eq:Vertices}
H_{SC}&=&\Delta^{(\ell)}_{m;\mathrm{v}}i\sigma^y_{\sigma\sigma'}d^\dagger_{\ell+m,\mathrm{v},\sigma}d^\dagger_{-m,-\mathrm{v},\sigma'}+h.c.\\
H_{CDW}&=&\rho^{[\ell]}_{m;\mathrm{v}}d^\dagger_{\ell+m,-\mathrm{v},\sigma}d_{m,1+\mathrm{v},\sigma}\nonumber\\
H_{SDW}&=&\mathbf{M}^{[\ell]}_{m;\mathrm{v}}\cdot\boldsymbol{\sigma}_{\sigma\sigma'}d^\dagger_{\ell+m,-\mathrm{v},\sigma}d_{m,1+\mathrm{v},\sigma'}\nonumber
\end{eqnarray}
(summation over the indices is implied). \(\Delta^{(\ell)}_{m;\mathrm{v}}=\Delta^{(\ell)}(\mathbf{K}_{m,\mathrm{v}})\) is the SC/PDW vertex, while \(\rho^{[\ell]}_{m;\mathrm{v}}\), and \(\mathbf{M}^{[\ell]}_{m;\mathrm{v}}\) are the CDW, and SDW vertices with momentum transfers \((\pi,\pi)/q+\ell\mathbf{Q}\).
We use the notation \(M^{[\ell,j]}_{m;\mathrm{v}}\) to denote the \(j^{th}\) component of \(\mathbf{M}^{[\ell,j]}_{m;\mathrm{v}}\), including CDW as a special case with \(j=0\), \(\rho^{[\ell]}=M^{[\ell,0]}\)).

The vertex RG flow is obtained using the diagrams shown in Fig. \ref{fig:RGflowVertex}, which yields
\begin{align}\label{RGvertexFlowEq}
\dot{\Delta}^{(\ell)}_{m;0}&=-g_{nm}^{(\ell)1}\Delta^{(\ell)}_{n;0}-g_{mn}^{(\ell)4*}\Delta^{(\ell)}_{n;1}\\
\dot{\Delta}^{(\ell)}_{m;1}&=-g_{nm}^{(\ell)4}\Delta^{(\ell)}_{n;0}-g_{nm}^{(\ell)1'}\Delta^{(\ell)}_{n;1}\nonumber\\
\dot{\rho}^{[\ell]}_{m;0}&=d_{ph}\left(g_{n-m,0}^{(\ell+m-n)2}-2g_{0,-\ell}^{(\ell-m+n)3}\right)\rho^{[\ell]}_{n;0}+d_{ph}\left(g_{0,n-m-1}^{(\ell+m-n)4*}-2g_{0,-\ell}^{(\ell+m-n)4*}\right)\rho^{[\ell]}_{n;1}\nonumber\\
\dot{\rho}^{[\ell]}_{m;1}&=d_{ph}\left(g_{n-m,-1}^{(\ell+m-n)4}-2g_{0,-\ell}^{(\ell-m+n)4}\right)\rho^{[\ell]}_{n;0}+d_{ph}\left(g_{1-\ell,1-\ell+n-m}^{(\ell+m-n-1)2}-2g_{1-\ell,0}^{(\ell+m-n-1)3}\right)\rho^{[\ell]}_{n;1}\nonumber\\
\dot{M}^{[\ell]}_{m;0}&=d_{ph}\left(g_{n-m,0}^{(\ell+m-n)2}M^{[\ell]}_{n;0}+g_{0,n-m-1}^{(\ell+m-n)4*}M^{[\ell]}_{n;1}\right)\nonumber\\
\dot{M}^{[\ell]}_{m;1}&=d_{ph}\left(g_{n-m,-1}^{(\ell+m-n)4}M^{[\ell]}_{n;0}+g_{1-\ell,1-\ell+n-m}^{(\ell+m-n-1)2}M^{[\ell]}_{n;1}\right)
\,
\nonumber
\end{align}

In order to identify the leading instability, we consider the susceptibilities \(\chi_I\) with \(I=\Delta^{(\ell)}_{m;\mathrm{v}},\, \tilde{\rho}^{[\ell]}_{k;\mathrm{v}},\, \tilde{M}^{[\ell]}_{k;\mathrm{v}}\). The susceptibilities flow as \(\dot{\chi}_I=d_I \left|I(t)/I(0)\right|^2\) with \(d_\Delta=1\) and \(d_I=d_{ph}\), and with \(\chi_I(0)=0\). The fastest diverging susceptibility corresponds to the leading instability. Solving the flow equations numerically we obtain the phase diagram Fig. \ref{fig:Phases} (b). See \cite{ShafferWangSantos22} for additional details.

\end{document}